\pdfoutput=1
\documentclass[11pt]{article}
\usepackage[affil-it]{authblk} 
\usepackage[authoryear]{natbib}
\usepackage{gensymb}
\usepackage{amsmath}
\usepackage{amssymb}
\usepackage{graphicx}
\usepackage{float}
\usepackage{booktabs}
\usepackage[a4paper, top=1in,bottom=1in,left=1in,right=1in,marginparwidth=0cm]{geometry}
\usepackage[title]{appendix}
\usepackage{pdflscape}
\usepackage{longtable}

\usepackage{adjustbox}
\usepackage{array}
\usepackage{caption}
\usepackage{chngcntr}
\usepackage{mathtools}
\usepackage{lineno}
\usepackage{setspace}
\usepackage{makecell}
\usepackage{hyperref}
\hypersetup{colorlinks=true,allcolors=blue}

\newcolumntype{R}[2]{%
    >{\adjustbox{angle=#1,lap=\width-(#2)}\bgroup}%
    l%
    <{\egroup}%
}

\renewcommand{\eqref}[1]{Eq.\,\ref{#1}}

\title{Determinism vs. stochasticity in competitive flour beetle communities}
\author[1,*]{Evan C. Johnson}
\author[2]{Tad Dallas}
\author[3,4]{Alan Hastings}
\affil[1]{Mathematical and Statistical Sciences; University of Alberta; Edmonton, Alberta, Canada}
\affil[2]{Biological Sciences, University of South Carolina; Columbia, South Carolina, United States }
\affil[3]{Center for Population Biology; University of California Davis; Davis, California, United States}
\affil[4]{Santa Fe Institute, Santa Fe, New Mexico, United States}
\affil[*]{Corresponding author: Evan Johnson, ecjohns1@ualberta.ca}

\date{}

\begin{document}

\maketitle 

\newpage

\tableofcontents

\newpage 

\section*{Abstract}

\begin{enumerate}
\item As ecologists increasingly adopt stochastic models over deterministic ones, the question arises: when is this a positive development and when is this an unnecessary complication? While deterministic models --- like the Lotka-Volterra model --- provide straightforward predictions about competitive outcomes, they are often unrealistic. Stochastic models are more realistic, but their complexity can limit their usefulness in explaining coexistence.

\item Here, we investigate the relative importance of deterministic and stochastic processes in competition between two flour beetle species, \textit{Tribolium castaneum} and \textit{Tribolium confusum}. Specifically, we use highly-replicated one-generation experiments ($784$ microcosms) to parameterize a mechanistic model.

\item Both the full stochastic model and the underlying ``deterministic skeleton'' exhibit \textit{priority effects}, where one species excludes the other, but the identity of the winning species depends on initial abundances. Stochasticity makes the identity of the winner less predictable, but deterministic dynamics still make reliable predictions (94\% accuracy across a range of reasonable initial abundances). 

\item We conclude that deterministic population dynamics are sufficient to account for patterns of coexistence (or lack thereof), a potentially general finding that is supported by recent field studies. 

\item Additionally, we resolve longstanding issues in flour beetle research by identifying selective egg predation as the mechanism for priority effects, demonstrating the primacy of demographic stochasticity (compared to genetic drift), and reinterpreting classic competition experiments to show that apparent coexistence often represents long-term transient dynamics.

\end{enumerate}

\textbf{Keywords}: Flour beetle, competition, coexistence, Tribolium, demographic stochasticity, environmental stochasticity, determinism, population dynamics

\newpage

\section{Introduction} \label{Introduction}

The field of community ecology has evolved from relying on deterministic models to embracing stochastic models \citealp{coulson2004skeletons}; \citealp[p. 1]{rohde2006nonequilibrium}; \citealp{boettiger2018noise}). Deterministic models have the advantage of clearly linking competitive outcomes to demographic parameters. Most famously, the two-species Lotka-Volterra model contains competition coefficients, whose values lead to four possible outcomes: species \#1 wins, species \#2 wins, both species coexist, or the phenomenon known as \textit{priority effects}, where neither species can invade its competitor at equilibrium \citep{volterra1926variationsST, Lotka1932TheSupply}. While this theory is elegant, it is perhaps too simple.

Ecologists continue to deepen our understanding of how stochasticity (or ``noise'') influences population dynamics. The concept of stochasticity encapsulates the seemingly random fluctuations in population size due to environmental fluctuations, genetic differences, and plain luck. It typically represents epistemic, rather than ontological, uncertainty; i.e., our lack of knowledge about the detailed causes of population growth. Stochasticity can alter mean growth rates through a phenomenon known as \textit{nonlinear averaging} or \textit{Jensen's inequality}, potentially changing competitive outcomes or promoting coexistence \citep{armstrong1976coexistence, chesson1982stabilizing, chesson1994multispecies, boettiger2018noise}. Stochasticity can also cause superior competitors to go extinct, a phenomenon known as competitive indeterminacy \citep{park1962beetles, mertz1976experimental, griesemer1988causal}. Although less pertinent to the focal issue of long-term coexistence, stochasticity can explain short-term patterns including species turnover and abundance distributions \citep{kalyuzhny2015neutral, grilli2020macroecological}. Nonlinear population dynamics and stochasticity can interact synergistically, giving rise to population cycles \citep{nisbet1976simple, leirs1997stochastic}, synchronized population cycles \citep{grenfell1998noise}, chaos \citep{turchin2000living, beninca2015species}, episodic outbreaks \citep{hastings2021effects}, and intermittent stability \citep{moskalenko2018characteristics}.

The move towards stochastic modeling invites scrutiny. On one hand, perhaps the contemporary focus on stochasticity is the sign of a maturing field or the increasing availability of ecological data. On the other hand, perhaps the pendulum has swung too far in the other direction; perhaps stochastic models introduce unnecessary complexity without fundamentally altering predictions.  Recent syntheses imply that both determinism and stochasticity matter \citep{vellend2010conceptual, barabas2018chesson, leibold2004metacommunity}. While strictly true, the ``everything is important'' perspective is not very enlightening. Given the overwhelming complexity of ecosystems, finding the most important factors is crucial to understanding. Parsimonious models in ecology serve several purposes: 1) making generalizations at the cost of precision \citep{levins1966strategy} 2) improving predictive performance \citep{shmueli2010explain}, and 3) satisfying aesthetic preferences \citep[p. 190]{kingsland1995modeling}. Here, our desire for parsimony is based on a fourth purpose: simple models are easier to write and analyze, making them a more convenient vehicle for understanding and scientific explanation \citep{keuzenkamp2002enigma}.

Does adding stochasticity to ecological models improve one's understanding of competitive outcomes? Answering this question is challenging due to the complexity of modeling real ecosystems. Studies of coexistence in field populations often utilize simple models (e.g., Ricker, Beverton-holt, Lottery model, or Annual plant model) whose ability to capture real-world complexity is questionable. More often than not, authors provide no justification for model structure. A potentially better approach is to utilize microcosm experiments. Microcosms are simple enough to be well-understood, and while somewhat unrealistic, remain more grounded than purely theoretical models \citep{lawton1996can}. Microcosms trade external validity (generalizing to the real world) for internal validity (understanding the system correctly). The loss of external validity can be managed by recognizing the unrealistic aspects of microcosms and using previous research to understand potential biases (see the \textit{Discussion}, Section \ref{Discussion}).

Here, we use flour beetle microcosms to study the relative importance of deterministic and stochastic processes in determining ecological coexistence. Specifically, we performed one-generation experiments with two species (\textit{Tribolium castaneum} and \textit{Tribolium confusum}), parameterized competition models, and then simulated long-term competitive outcomes under various scenarios (e.g., turning off demographic stochasticity). Our study aims to contribute to a broader discussion regarding the level of detail required to accurately capture ecological coexistence.

\section{Methods} \label{Methods}

\subsection{Tribolium competition experiment} \label{Tribolium competition experiment}

Two species of flour beetles, \textit{Tribolium confusum} \& \textit{Tribolium castaneum}, were obtained from long-running laboratory stock populations. Both stock populations and experimental populations faced similar environments: plastic enclosures with dimensions 4 x 4 x 6 cm, filled with 30 ml of ``standard media'', consisting of 95\% wheat flour and 5\% brewer's yeast by weight. All populations were maintained at 30°C and approximately 50\% relative humidity. 

To simplify population dynamics, we imposed non-overlapping generations, increasing the generality of our system for temperate-zone insects with synchronized life cycles. Adult beetles were placed in enclosures with fresh media for 24 hours, during which they oviposited and consumed eggs of both species. This phase is called the \textit{oviposition period}. Afterward, the adults were removed, and the eggs were allowed 5 weeks (minus the oviposition period) to develop into adults. This protocol minimizes larvae eating eggs; this inter-stage interaction occurs at an instar-specific rate \citep{hastings1987cannibalistic, hastings1991oscillations}, and is thus difficult to model. Other inter-stage interactions, specifically adults and larvae eating pupae, are considered negligible \citep{park1965cannibalistic, hagstrum1980age}. After the 5-week developmental period, we identified and counted the adult beetles by species.

The initial abundances varied from 0 to 256 adults of each species per enclosure (0, 2, 4, 8, 16, 32, 64, 128, 256). We considered all (non-zero) two-species combinations for abundances between 0 and 64, with 15 replicates per initial abundance combination. Additionally, all (non-zero) two-species combinations of 0, 128, and 256 were replicated 8 times each. In total, the experiment contained 768 replicates/enclosures. One replicate containing 64 \textit{T. castaneum} and 0 \textit{T. confusum} was discarded due to experimental error.

\subsection{Stochastic Model}

\textit{T. confusum} and \textit{T. castaneum} are respectively denoted with subscripts ``1'' \& ``2''. For simplicity, we only present equations for \textit{T. confusum}, noting that the equations for \textit{T. castaneum} are obtained simply by swapping subscripts. 

Upon introducing adults into the experimental patches, the number of females is unknown, as reliable sexing is only possible at the pupal stage. Thus, we treat the number of females as a binomial random variable,
\begin{equation} \label{eq:female_sample}
    f_1(t) \sim \text{Binomial}\left(n_1(t), \text{prob} = 0.5 \right),
\end{equation}
where $f_1(t)$ is the number of females in generation $t$; $n_1(t)$ is the number of initial \textit{T. confusum} adults, and the 50\% ``success'' probability reflects the expected proportion of females \citep{howe1956effect}. 

At the start of each generation, the females oviposit at an average rate of $\alpha_1$ eggs per female per day. However, the actual number of oviposited eggs per female fluctuates with standard deviation $\eta_1$. Females of \textit{T. confusum} adult females cannibalize their eggs at the rate $\beta_{11}$, whereas \textit{T. castaneum} adult females eat \textit{T. confusum} eggs at the rate $\beta_{12}$. A previous study found that oviposition and cannibalism rates are approximately constant for at least 48 hours \citep{rich1956egg}. We assume that males do not eat eggs, as \textit{Tribolium} females are the more voracious sex \citep{boyce1946influence, stanley1942mathematical, rich1956egg}, with one experiment finding that females eat 19 times as many eggs as males \citep{sonleitner1961factors}. All of the above processes can be captured in the stochastic differential equation (SDE),

\begin{equation} \label{eq:sde}
    dz_1(h) = \left[ f_1(t) \alpha_1 - \beta_{11} z_1(h) f_1(t)  - \beta_{12} z_1(h) f_2(t) \right] dh + \sqrt{f_1(t)} \eta_1 dW(h),
\end{equation}
where $z$ is the number of eggs, $h$ is the number of days since the beginning of the oviposition period, and $dW(h)$ is an increment of the Weiner process \citep{karlin1981second}. Each oviposition period starts with fresh media, so the initial condition is $z_1(0) = 0$.

For simplicity, the model assumes that all eggs at the end of the oviposition period survive to adulthood. The egg-to-adult survival probability is around 90\% under ideal environmental conditions \citep{sokoloff1974biology, johnson2022explanation}, but explicitly modeling this process would create a 3-layer hierarchical model with computational challenges; note that the binomial sampling of females means that the model is already hierarchical. The average and variance in developmental survival are ``absorbed'' in the model-fitting process by $\alpha_1$ and $\eta_1$ respectively. 

The SDE for eggs (\eqref{eq:sde}) can be converted to its Fokker-Planck representation and solved with the Fourier method. The result is a multivariate normal distribution where the expected number of \textit{T. confusum} adults in the following generation, assuming an oviposition period of $s$ days, is
\begin{equation}
    \mu_1(f_1, f_2) = \frac{f_1 \alpha_1 \left(1-\exp \left[-s \left( \beta_{11} f_1 + \beta_{12} f_2 \right) \right] \right)}{\beta_{11} f_1 + \beta_{12} f_2}.
\end{equation}
The variance in the number of \textit{T. confusum} adults is
\begin{equation}
   \sigma_1^2(f_1, f_2) = \frac{f_1 \eta_1^2 \left(1-\exp \left[-2 s \left( \beta_{11} f_1 + \beta_{12} f_2 \right) \right] \right)}{2\left( \beta_{11} f_1 + \beta_{12} f_2\right)},
\end{equation}
and there is no covariance between the net fecundity of the two species, due to the that assumption the fluctuations in oviposition rates are independent.

The stochastic differential equation generates a normal distribution for the density of beetle progeny, which is unrealistic in two significant ways. First, it is a continuous distribution, whereas we observe a discrete number of adults. To fix this, we apply a continuity correction: the probability of observing $x$ individuals is the integral of the probability density function (PDF) from $x-0.5$ to $x+0.5$. Second, the normal distribution has support on negative values, whereas the least number of observable adults is zero. To fix this, we add \textit{zero-inflation} to the model, such that the probability of observing zero individuals is the integral of the PDF $-\infty$ to $0.5$. These modifications are validated by leave-one-out cross-validation \cite{vehtari2019loo} --- the zero-truncated normal distribution with a continuity correction outperformed both the zero-truncated normal and reparameterized negative binomial distributions (see the \textit{Supporting materials}, Section \ref{Data and code availability statement}). A random variate of the modified normal distribution, denoted with an asterisk *, gives the number of adults in the following generation:

\begin{equation} \label{eq:adult_sample}
    n_1(t+1) \sim \text{Normal}^*\left(\mu_1(f_1(t), f_2(t)), \sigma_1^2(f_1(t), f_2(t)) \right).
\end{equation}

All models were fit using the program \textit{Stan} \citep{stan2020rstan}, which implements Hamiltonian Monte Carlo. We monitored a standard suite of model-fitting diagnostics to reassure ourselves that estimation was unbiased and that the Markov chains had converged to the global posterior distribution, specifically $\hat{R}$, divergences, effective sample size, and the Bayesian fraction of missing information\citep[ch. 6]{gelman2014bayesian}. We utilized weakly informative priors, and computed the posterior contraction \citep{schad2021toward} to ensure that the priors did not have a significant effect on the results.

\subsection{Model justification} \label{Stochasticity}

Four types of stochasticity could potentially affect population dynamics. 1) Demographic stochasticity \citep{kendall2003unstructured, lande2003stochastic, may1973stability} stems from moment-to-moment variation in individuals' demographic rates. The effect of demographic stochasticity on population growth rates increases linearly with $\sqrt{n}$; this scaling is essentially the central limit theorem for a sum of i.i.d. random variables. 2) Environmental stochasticity \citep{lande2003stochastic} is the randomness in the population-level average of a demographic parameter, due to environmental fluctuations. Because individuals within a species have correlated responses to the environment, the effect of environmental stochasticity on population growth increases linearly with population density $n$. 3) Demographic heterogeneity \citep{kendall2003unstructured, melbourne2008extinction} is inter-individual variation in demographic rates, due to persistent differences in traits or behaviors. 4) Sex-ratio stochasticity \citep{engen2003demographic, lande2003stochastic} is the deviation from the expected sex ratio due to Mendel's law of segregation. Sex-ratio stochasticity can be thought of as a particular form of demographic heterogeneity (\#3 above), where the demographic rate in question is fecundity, and the trait in question is biological sex. To summarize, environmental stochasticity and sex-ratio stochasticity are self-explanatory, demographic heterogeneity stems from trait variation, and demographic stochasticity is luck.

In our flour beetle microcosms, the predominant forms of stochasticity are demographic stochasticity and sex-ratio stochasticity. The full justification of this statement, which involves graphical evidence and extensive model comparisons, is given in Appendix S4 of \citet{johnson2022explanation}. We summarize the evidence here. Environmental stochasticity is negligible, as evidenced by a sublinear scaling between $sd(n_2(t+1))$ and $n_2(t)$, and the fact that we attempted to hold the environment constant (though we cannot definitively rule out the presence of incubator microclimates). Demographic heterogeneity is the only way to account for the high variance in $n_2(t+1)$ at small $n_2(t)$ for experiments with a long, 7-day oviposition period: between-individual differences in fecundity have compounding effects on egg production over the oviposition period (because the sex ratio stays the same), whereas demographic stochasticity does not (a large number of eggs one day is likely to be canceled out by a small number of eggs the next day). Based on the fact that sex-ratio stochasticity undeniably exists and that models with additional heterogeneity do not boost predictive performance, we determined that sex-ratio stochasticity was the dominant form of demographic heterogeneity in our system; recall here that sex-ratio stochasticity is a special case of demographic heterogeneity. Last but not least, demographic stochasticity was consistently present in the top-performing models, and was able to explain the persistence of variation in large populations: in a model with only demographic heterogeneity, variation in $n(t+1)$ goes to zero as $n(t)$ goes to infinity.

\subsection{Stochastic simulations}

The stochastic model can easily be simulated by selecting initial abundances and iteratively applying the equations, \eqref{eq:female_sample}--\eqref{eq:adult_sample}. To better understand the role of stochasticity in competition, we simulate the model under counterfactual scenarios where various types of stochasticity are present/absent. To turn off sex-ratio stochasticity, we simply replace the binomial sampling of females (\eqref{eq:female_sample}) with the expectation $f_1(t) = n_1(t)/2$; to turn off demographic stochasticity, we fix the noise parameters at zero, i.e.,  $\eta_1 = \eta_2 = 0$. To attain the \textit{deterministic skeleton}, an idealized representation that is most amenable to mathematical analysis, we turn off sex-ratio stochasticity, demographic stochasticity, and the continuity correction, such that the $n_1(t+1)$ can take non-integer values.

\subsection{Coexistence analysis} \label{Coexistence analysis} 

Using classic ecological theory, we can derive the conditions for coexistence in the deterministic version of our flour beetle model. This exercise builds intuition about how coexistence is related to flour beetle demography. Moreover, coexistence outcomes in the deterministic skeleton serve as a baseline for understanding the effects of stochasticity.

The usual methodology for studying coexistence is an \textit{invasion analysis}  \citep{turelli1978reexamination, Grainger2019TheResearch}, where we ask: if introduced at low density, can a species invade a community of its heterospecific competitors at their equilibrium? In a two-species community without positive density dependence, coexistence is implied if both species can invade \citep{chesson1989invasibility}; competitive exclusion is implied if only one species can invade; and the phenomenon known as \textit{priority effects} or founder control \citep{fukami2015historical} is implied if neither species is able to invade.

First, consider the deterministic time-1 map for species $i$:
\begin{equation} \label{eq:pop_map_det}
    n_i(t+1) = \frac{\frac{n_i}{2} \alpha_i \left(1-\exp \left[-s \left( \beta_{ii} \frac{n_i}{2} + \beta_{ij} \frac{n_j}{2} \right) \right] \right)}{\beta_{ii} \frac{n_i}{2} + \beta_{ij} \frac{n_j}{2}}. 
\end{equation}
Next, divide both sides of the equation to obtain the per capita replacement rate (also known as the finite rate of increase; \citealp{gotelli1995primer}); set the focal species density at $n_i = 0$ to approximate low-density conditions; and set the heterospecific to its equilibrium density, $n_j^*$. The low-density finite rate of increase can then be written as 
\begin{equation}
    \lambda_{i}^{\{-i\}} = \frac{\alpha_i \left(1-\exp \left[-s \beta_{ij} \frac{n_j^*}{2} \right] \right)}{2 \beta_{ij} \frac{n_j^*}{2}}.
\end{equation}

The invasibility condition for coexistence is $\lambda_{1}^{\{-1\}} > 1$ \& $\lambda_{2}^{\{-2\}} > 1$, where the superscript is the index of the species at low density. Expanding $\lambda_{i}^{\{-i\}}$ is not so insightful, since the equilibrium density does not have a closed-form solution:
\begin{equation}
    n_j^* = \frac{s \alpha_j + 2 \, \Omega\left(-\frac{1}{2} \, s \, \alpha_j \, \exp\left[- \frac{s \, \alpha_j}{2} \right]\right)}{s\,\beta_{jj}},
\end{equation}
with $\Omega$ representing the Lambert $W$ function, the inverse function of $f(W) = We^W$ \citep{lehtonen2016lambert}. However, in the limit of large $s_j \, \alpha_j$, the equilibrium density becomes $n_j^* = \alpha_j / \beta_{jj}$, and the low-density finite rate of increase becomes $\lambda_{i}^{\{-i\}} = (\alpha_i \beta_{jj}) / (\alpha_j \beta_{ij})$. 

With the above simplification, the condition for coexistence becomes
\begin{equation}
    \frac{\beta_{11}}{\beta_{21}} > \frac{\alpha_1}{\alpha_2} > \frac{\beta_{12}}{\beta_{22}}, 
\end{equation}
which is structurally identical to the condition for coexistence in the two-species Lotka-Volterra model \citep[pg. 136]{hastings1997population}. To be clear, our analysis of coexistence for the deterministic skeleton uses the exact invasion growth rates, $\lambda_{1}^{\{-1\}}$ \& $\lambda_{2}^{\{-2\}}$. However, the large $s_j \alpha_j$ approximation above demonstrates that simpler models are similar to our model, with the additional complexity of our model arising from transient dynamics. The total oviposition is proportional to the number of females, but egg predation is proportional to the product of females and eggs. Therefore, it takes time for the egg count to reach a level where oviposition and egg predation reach equilibrium.

\section{Results} \label{Results}

\begin{figure}[H]
\centering
\makebox{\includegraphics[scale = 1]{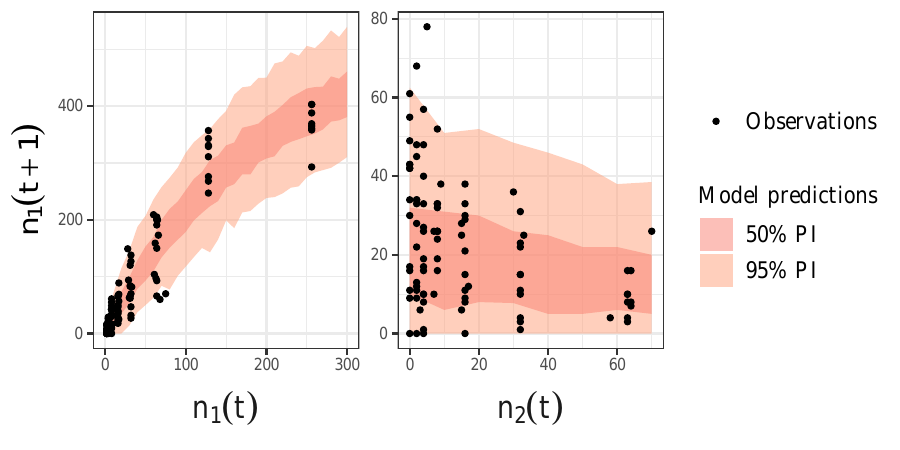}}
\caption{The statistical model makes predictions that match experimental observations. The salmon ribbons show the predictive intervals (PI) of simulated data. The left panel was generated by fixing \textit{T. castaneum} initial abundance at $n_2(t) =$ 0, whereas the right panel was generated by fixing \textit{T. confusum} initial abundance at $n_1(t) =$ 8. }
\label{fig:pop_map_val}
\end{figure}

The parameterized model was able to successfully replicate empirical patterns in the data (Fig. \ref{fig:pop_map_val}). Further, point estimates of the oviposition rate parameters are consistent with previous research. We estimate that \textit{T. confusum}'s and \textit{T. castaneum}'s oviposition rates are respectively $\alpha_1 = 5.6$ and $\alpha_2 = 9.4$ eggs per female per day (Table \ref{tab:pars}); Previous research places \textit{T. confusum}'s and \textit{T. castaneum}'s oviposition rates at 4.4 and 8.6 eggs per female per day (\citealp{rich1956egg}, Table IV, mean of the ``mean column across treatments'' column; \citealp{sonleitner1961factors}, Table 2).  In part, our model is justified by the experimental setup: sifting out the adult beetles after a brief oviposition period results in non-overlapping generations, which justifies our discrete-time treatment and the omission of inter-stage interactions except for adults eating eggs. Previous flour beetle studies utilized the same experimental protocol but did not explicitly model the oviposition/egg-consumption dynamics. Instead, studies have used the Ricker model (e.g., \citealp{melbourne2008extinction}; \citealp{dallas2019can}) with the form $n(t+1) = n(t) R \exp(-\beta n(t))$, which incorrectly implies that each female immediately and instantaneously oviposits $R/2$ eggs. The Ricker model also implies a log-linear relationship between per capita growth rates and population abundance, $\log(n(t+1)/n(t)) = \log(R) - \beta n(t)$, whereas experimental data clearly show a curvilinear relationship (Fig. \ref{fig:growth_rates}). These modeling decisions can have a large impact on one's conclusions: the Ricker model implies the two flour beetle species should coexist \citep{dallas2021initial}, whereas our model implies priority effects.

\begin{table}[ht]
\centering
\begin{tabular}{lllll}
  \hline
Parameter & Mean & SD & $\text{CI}_{2.5\%}$ & $\text{CI}_{97.5\%}$ \\ 
  \hline
$\alpha_1$ & 5.63 & 0.187 & 5.28 & 5.99 \\ 
  $\alpha_2$ & 9.39 & 0.366 & 8.70 & 10.1 \\ 
  $\eta_1$ & 8.85 & 0.333 & 8.23 & 9.50 \\ 
  $\eta_2$ & 12.9 & 0.478 & 12.0 & 13.9 \\ 
  $\beta_{11}$ & 0.0107 & 0.00135 & 0.00817 & 0.0134 \\ 
  $\beta_{12}$ & 0.0377 & 0.00317 & 0.0316 & 0.0439 \\ 
  $\beta_{21}$ & 0.0270 & 0.00356 & 0.0201 & 0.0342 \\ 
  $\beta_{22}$ & 0.0428 & 0.00305 & 0.0371 & 0.0489 \\ 
   \hline
\end{tabular}
\caption{Model parameter estimates; includes posterior means, standard deviations, and the bounds of marginal 95\% credible intervals (CI). The parameters $\alpha_i$ are the oviposition rates, units: eggs $\cdot$ female$^{-1}$ $\cdot$ day$^{-1}$; $\eta_i$ are the scales of demographic stochasticity in total oviposition, units: eggs $\cdot$ female$^{-1/2}$ $\cdot$ day$^{-1/2}$; and $\beta_{ij}$ are the egg-eating rates, units: eggs $\cdot$ female$^{-1}$ $\cdot$ eggs$^{-1}$ $\cdot$  day$^{-1}$. }
\label{tab:pars}
\end{table}

The deterministic dynamics give rise to priority effects. Using an invasion analysis (Section \ref{Coexistence analysis}), we see priority effects (i.e., both species have negative invasion growth rates) for 97\% of draws from the posterior distribution of model parameters. Because there was little uncertainty with respect to the deterministic competitive outcomes, and because of generally low levels of posterior variation, we use the posterior means of model parameters for the stochastic simulations.

\begin{figure}[H]
\centering
\makebox{\includegraphics[scale = 0.8]{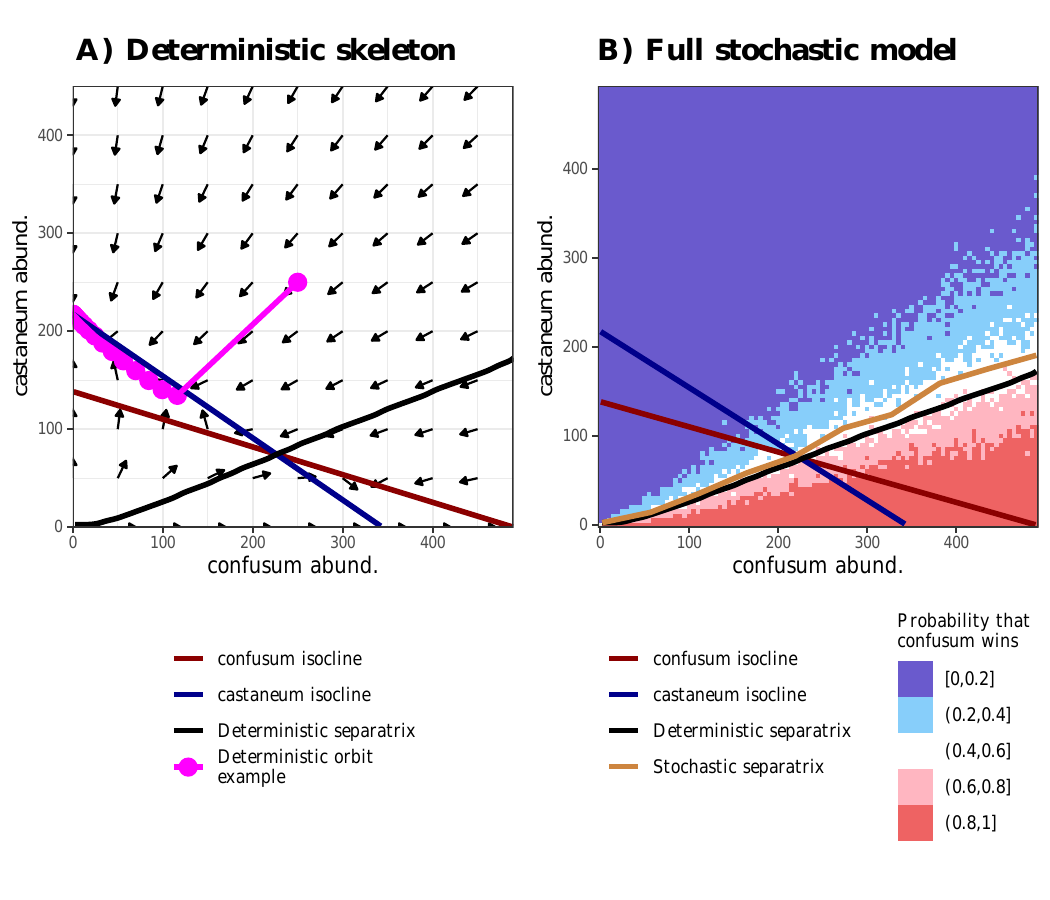}}
\caption{Two-species phase plane diagrams for A) the deterministic skeleton, and B) the full stochastic model. The introduction of stochasticity introduces competitive indeterminacy, as shown in the light-color funnel in Panel B. Additionally, stochasticity slightly favors \textit{T. confusum} on average; to see this, compare the deterministic and stochastic separatrices in panel B.}
\label{fig:phase_planes}
\end{figure}

The stochastic dynamics also show the telltale signs of priority effects (Fig. \ref{fig:phase_planes}): both species cannot coexist in the long run, and the likely winner depends on initial abundance. However, stochasticity alters the dynamics in several ways. The most prominent impact is \textit{competitive indeterminacy}, where the winner of competition is somewhat unpredictable, even when initial abundances are known. The degree of competitive indeterminacy is context-dependent. If the initial abundances start near the \textit{separatrix} --- a line of initial abundances that separates the basins of attraction for either species' stable equilibria \citep[pg. 160]{strogatz2018nonlinear} --- then each species has a roughly 50\% chance of winning. If the initial conditions give one species a much higher abundance, then unsurprisingly, that species is more likely to win. The degree of indeterminacy also increases as both species' initial abundances increase, owing to the fact that demographic stochasticity scales with population size; this is evidenced by the funnel-shape of the white panels, denoting roughly equal chances of competitive exclusion, in the stochastic phase plane diagram (Fig \ref{fig:phase_planes}, panel B).

\begin{figure}[H]
\centering
\makebox{\includegraphics[scale = 1]{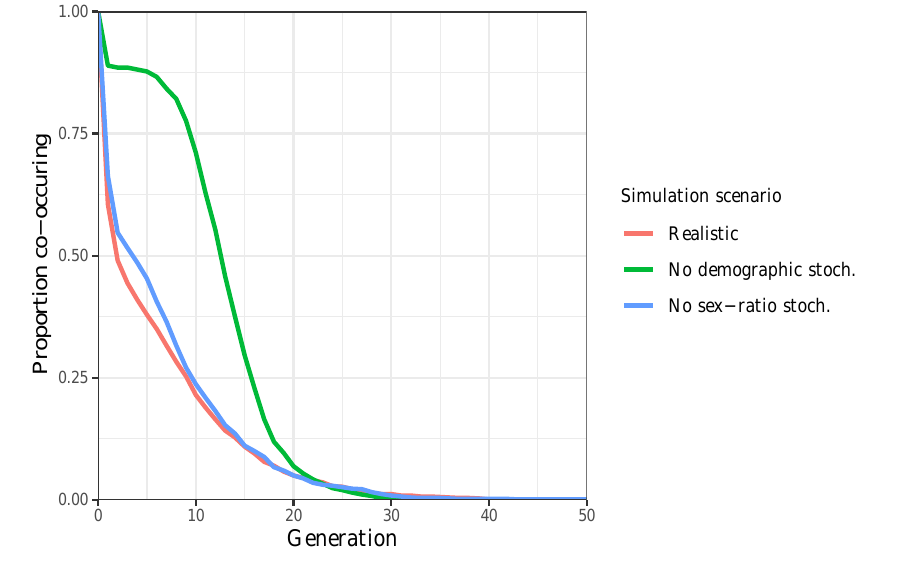}}
\caption{Demographic stochasticity shortens the length of transient co-occurrence. We performed simulations with initial conditions of 4 adults per species, recording the proportion of simulations with co-occurrence, i.e., neither species had yet been competitively excluded. In the realistic simulation scenario (i.e., the full model), and in the ``no sex-ratio stochasticity'' scenario, only 25\% of the simulations had co-occurrence by the tenth generation. When demographic stochasticity was turned off, 75\% of the simulations exhibited co-occurrence by the tenth generation.}
\label{fig:co_occur_remaining}
\end{figure}

Stochasticity accelerates competitive exclusion (Fig. \ref{fig:co_occur_remaining}). Without stochasticity, the two species co-occur for more generations, consistent with previous research (e.g., \citealp{carmel2017using}; \citealp{lande1998demographic}). This is because the long-term growth rate is the geometric mean of $\lambda$, approximated by the arithmetic mean minus half the temporal variation in $\lambda$ \citep{Lewontin1969}. In other words, temporal variation in the generational growth rate inherently depresses the relevant long-term growth rate.

Demographic stochasticity plays a much larger role than sex-ratio stochasticity. The size of the funnel of competitive indeterminacy in the stochastic phase plane diagram shrinks significantly in the absence of demographic stochasticity, but not in the absence of sex-ratio stochasticity (Fig. \ref{fig:stoch_phase_scenarios}). Further, transient co-occurrence is much longer in the absence of demographic stochasticity, but not in the absence of sex-ratio stochasticity.

\section{Discussion} \label{Discussion}

Are stochastic population dynamics necessary to understand patterns of coexistence? In our experimental flour beetle communities, the answer is ``No''. Both the full stochastic model and the underlying deterministic skeleton lead to patterns of priority effects, where the winner of competition depends on initial abundances. Thus, stochasticity does not qualitatively change the competitive dynamics. The inclusion of stochasticity leads to some degree of competitive indeterminacy, but the outcome of competition is highly predictable for most initial abundances.

The effect of stochasticity is context-dependent, in the sense that there is a great deal of uncertainty about which species will win if initial abundances start near the separatrix, and not so much uncertainty otherwise. There is no single quantitative measure of the ``importance of stochasticity''. Visual inspection of the stochastic phase plane diagram (Fig. \ref{fig:phase_planes}) shows that there is a relatively narrow band of initial abundances where the two species have roughly equal chances --- at the unstable equilibrium point, a 40-60\% probability of winning is attained with roughly +/- 10 beetles of either species. If we assume initial conditions selected at random between each single-species equilibrium (i.e., the cartesian product of [1, $n_1^{*\{-2\}}$] and [1, $n_2^*{\{-1\}}$]), the deterministic skeleton accurately predicts the winner of competition 94\% of the time. The uncertainty is almost entirely due to demographic stochasticity, since the prediction accuracy rises to 99\% for a hypothetical community where demographic stochasticity has been turned off. 

Our case study of flour beetles suggests that stochasticity isn't crucial for understanding coexistence, an idea which finds some support in recent research. One way in which stochasticity can affect coexistence is the \textit{storage effect}. This fluctuation-dependent coexistence mechanism, wherein species specialize in different states of a variable environment, is thought to be the explanation for how many species can coexist on a few resources (i.e., the paradox of the plankton; \citealp{hutchinson1961paradox}). However, \citet{stump2023reexamining} estimate that the storage effect is weaker than classic coexistence mechanisms (e.g., resource partitioning), even in communities where the storage effect is believed to occur, specifically tropical trees of Barro Colorado Island, and communities of annual plants in Spain, California, and Mexico \citep{godoy2014phenology, hallett2017functional, zepeda2019fluctuation}. The storage effect arises naturally when fecundity fluctuates in a temporally autocorrelated fashion \citep{schreiber2021positively}, but most animals have temporally uncorrelated or weakly autocorrelated demographic rates \citep{knape2011effects, ferguson2016updated}.

Additional research supports the sufficiency of deterministic dynamics in predicting competitive outcomes. Simulations of annual plants in serpentine hummocks demonstrate that deterministic population dynamics accurately predict coexistence for thousands of years unless initial abundances were extremely low \citep{schreiber2023does}. Another fluctuation-dependent coexistence mechanism known as relative nonlinearity \citep{armstrong1976coexistence, chesson1994multispecies}, is unlikely to be a general explanation for multi-species coexistence due to its limitation of supporting only two species per discrete resource \citep{johnson2023coexistence}. Consequently, most empirical studies have focused on quantifying the storage effect rather than relative nonlinearity (e.g., \citealp{descamps2005stable, angert2009functional}). Some studies have found significant contributions of relative nonlinearity to invasion growth rates \citep{zepeda2019fluctuation, letten2018species}. Importantly, in these study systems, relative nonlinearity stems from environmental stochasticity, and therefore does not permit multi-species coexistence without additional coexistence mechanisms (\citealp{stump2017optimally}, Appendix D.2).


To be clear, we don't aim to reopen the debate on stochasticity's importance in ecology; this has been settled. In the 1930s, W.R. Thompson challenged mathematical models in ecology, arguing natural phenomena were too complex for equations to predict \citep[pg. 140]{kingsland1995modeling}. While Thompson's view was ultimately rejected, his admonition contributed to the incorporation of stochasticity in ecological models (e.g., \citealp{bartlett1960stochastic}). By the 1960s, stochastic models were widely used, though interest fluctuated: equilibrium-based approaches were popularized by Robert Macarthur in the 1970s \citep{kingsland1995modeling}, followed by a stochasticity resurgence around 2000 with Neutral Theory \citep{hubbell2001unified} and Modern Coexistence Theory \citep{chesson2000mechanisms}. Undoubtedly, stochasticity matters, especially in community assembly, epidemiology, and biological invasions. Our claim is more modest: deterministic models can often account for patterns of long-term coexistence. 

Because microcosms are deliberately unrealistic, it is worth considering how the complexities of the real world may alter our results. Our microcosms have a small spatial extent and lack environmental variation. Environmental variation can promote coexistence via the storage effect; in our microcosms, this would require that temperature or relative humidity be autocorrelated on the timescale of multiple generations \citep{johnson2022towards}. A larger spatial extent would lessen the effect of demographic stochasticity on the overall population density, as patches with many eggs are ``canceled out'' by patches with few eggs (i.e., \textit{statistical stabilization}, \textit{sensu }\citealp{de1991mobility}). Thus, a larger spatial extent would only reinforce our main finding that deterministic dynamics predict long-term coexistence. 

Our work is situated within a rich literature on \textit{Tribolium} competition, providing new insights and resolving existing issues. Thomas Park and his collaborators conducted long-term experiments of competition between \textit{T. confusum} and \textit{T. castaneum}, resulting in several influential publications. The earliest paper \citep{park1948interspecies}, now canalized in \textit{Foundations of Ecology} textbook \citep{real1991foundations}, found that one species was always excluded by the other, but that the identity of excluded species could vary even under identical experimental conditions, thereby introducing the concept of \textit{competitive indeterminacy}. In the second competition paper \citep{park1954experimental}, varying environmental conditions (i.e., incubator temperature and humidity) revealed that different environments favored different species, with no intermediate state allowing coexistence. In the third paper \citep{park1957experimental}, varying initial abundances under conditions favoring \textit{T. castaneum} showed \textit{T. confusum} winning only occasionally when starting at a severe numerical advantage. Taken as a whole, Park et al.'s experiments embraced both determinism and stochasticity. The relationships between environmental conditions (e.g., temperature) and demographic parameters (e.g., development time) were invoked to explain the dominance of one species in certain treatments, while ``chance cannibalism'' or ``inherent biological and statistical variability'' \citep{park1954experimental} were invoked to explain the unpredictability of competitive exclusion. Sadly, the original data from these long-term experiments were lost in a fire (Robert F. Costantino, personal communication).

The Park et al. experiments, where one species always competitively excluded the other, have often been interpreted as supporting the \textit{competitive exclusion principle}, the idea that no two species with an identical niche can coexist \citep{Gause1934, hardin1960competitive}. Our work demonstrates that despite inhabiting the same environment, the \textit{T. confusum} and \textit{T. castaneum} have distinct niches in the form of species-specific rates of oviposition and intraguild predation. It just so happens that the flour beetles' niches  (again, broadly construed in the sense of \citet{tilman1980resources};\citet{chase2003ecological}) actively undermine coexistence. Park et al. were more than aware of this possibility, but held a more physicalist interpretation of the ecological niche, writing that ``...it would follow that other strains of the two Tribolium might in time be found, which would exhibit the phenomenon of coexistence of both species in a competitive system. Theoretically speaking, the so-called Gause’s Principle does not necessarily hold in all possible cases.'' \citep{park1964genetic}. 

A number of previous papers have invoked priority effects to explain Park et al.'s experimental data \citep{barnett1962monte, leslie1962stochastic, mertz1976experimental, costantino1991population}. Our work reinforces this conclusion with a mechanistic model and highly-replicated experimental data. Too often, microcosms are treated purely as a means of removing complexity, e.g., removing environmental variation or multitrophic interactions. Here, we use microcosms to \textit{add} certain details --- temporally explicit egg dynamics, encoded in a solvable stochastic differential equation. It is an interesting coincidence that the first mathematical model of Tribolium competition, $\lambda / (1+ \alpha_1 n_1 + \beta_1 n_2)$ \citep{leslie1958properties}, is definitively phenomenological, yet is identical to the deterministic skeleton of our mechanistic model (\eqref{eq:pop_map_det}) in the limit of a long oviposition period.

Park et al.'s final competition experiment contained one replicate (out of 25) where both species co-occurred for the entire 960-day duration of the study \citep{leslie1968effect}. This single instance is intriguing, and has been explained as a case of long-term transient co-occurrence \citep{costantino1991population}, or as fluctuation-mediated coexistence \citep{edmunds2003park}, involving a mechanism known as relative nonlinearity \citep{barabas2018chesson}. Our analysis suggests that transient co-occurrence is the more likely explanation.

\begin{figure}[H]
\centering
\makebox{\includegraphics[scale = 1]{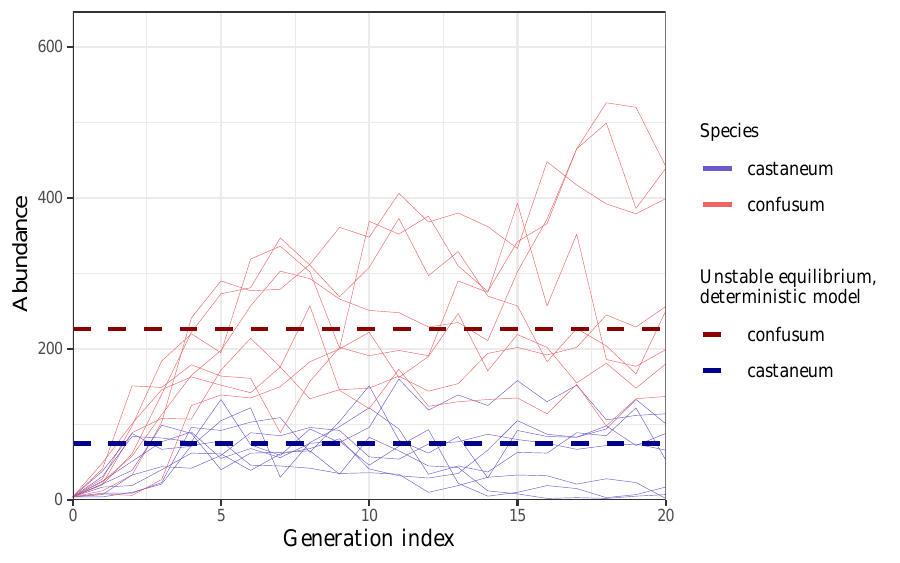}}
\caption{Due to selection bias, long-term transient co-occurrence can appear as fluctuations around a stable equilibrium, even though the equilibrium is actually unstable. Depicted are simulations where both species co-occurring for over 20 generations before castaneum goes extinct.}
\label{fig:sim_examples}
\end{figure}

Our simulations show that over 5\% of communities can coexist for more than 20 generations, equivalent to 700 days. While this is a significant duration, it suggests that the 960 days of co-occurrence observed in Leslie et al.'s (\citeyear{leslie1968effect}) experiment would be unlikely without stabilizing coexistence mechanisms. However, Leslie et al.'s experimental setup resulted in a small number of ``effective generations''. Adults were not removed from the flour, leading them to consume most of the eggs before they could hatch. Only when a significant number of adults died could a new cohort of eggs develop, subsequently causing a large and sudden increase in adult beetle numbers (Fig. 1 in \citealp{edmunds2003park}), sometimes referred to as \textit{single-generation cycles} \citep{murdoch2003consumer}. Thus, the speed of population dynamics is linked to the adult beetle's average lifespan, which can exceed a year \citep{sokoloff1974biology}. Leslie et al.'s data show relatively stable abundances over time, suggesting coexistence. However, this pattern can result from long-term transient co-occurrence due to selection bias. The pull of the single-species equilibria is weakest near the unstable two-species equilibrium, making transient co-occurrence often appear as fluctuations around a speciously stable equilibrium (Fig. \ref{fig:sim_examples}).

Due to the confidence we have in our model, we are also able to better identify demographic stochasticity and sex-ratio stochasticity as the dominant sources of randomness. Previously, it was uncertain whether competitive indeterminacy in Tribolium was caused by demographic stochasticity or a genetic founder effect \citep{lerner1961genotype, lerner1962indeterminism, dawson1970further, goodnight1996effect, mertz1976experimental}, which itself is a special case of demographic heterogeneity. The analysis performed by \citet{johnson2022explanation} (discussed further in Section \ref{Stochasticity}) demonstrates that there is no predictive benefit to including demographic heterogeneity in the system beyond sex-ratio stochasticity.

Environmental variation is almost entirely eliminated in our laboratory microcosms, which may raise concerns about generalizability --- environmental stochasticity is generally thought to have a larger impact because it increases faster with population density ($n$ vs. $\sqrt{n}$ for demographic stochasticity). However, research suggests that demographic and environmental stochasticity are comparable in significance within an order of magnitude at the spatial scales typical of most studies\citep{reed2004relationship, engen2001stochastic, saether1998environmental, saether2000population, saether2002density, snyder2022snared}. Thus, even with reduced environmental variation, the total impact of stochasticity in our experiments remains substantial. Conceptually, demographic stochasticity is large because it absorbs all individual-level variation that cannot otherwise be explained. Even genetic clones in laboratory conditions can exhibit large differences in life outcomes\citep{vogt2008production, bierbach2017behavioural, laskowski2022emergence}, putatively caused by a hopelessly intricate interaction between gene expression, microenvironment, and development.



Priority effects in Tribolium competition has previously been attributed to two main factors: discriminatory egg-eating and allelopathy. Flour beetles excrete secondary metabolites which accumulate in the flour, a process called ``conditioning''. Flour conditioning has both intra and interspecific effects \citep{park1936studies, prus1961effect,flinn2012effects, bullock2020interspecific}, reducing fecundity and slowing developmental times. There is a clear mechanism for stronger interspecific effects, which would be expected under priority effects: Because \textit{Tribolium} individuals are often in the vicinity of conspecifics, they have evolved to tolerate conspecific metabolites. Flour conditioning probably had a minimal impact in our experiments because the adult beetles were only exposed to the flour for 24 hours, while significant conditioning effects typically require several days to a week \citep{johnson2022explanation}. Flour conditioning typically kills early instar larvae, and we found few larval carcasses at census time. 

The priority effects in this study are due to selective egg-eating behavior. Unlike Pacman straightforwardly gulping down pellets, adult-on-egg predation in Tribolium is complex and costly, requiring specific leg positioning and up to 15 minutes to complete \citep{park1934observations}. This behavior may therefore be subject to evolutionary pressures. Females do not tunnel randomly; instead, they maintain home tunnel systems that change slowly over time \citep{rich1956egg, stanley1949mathematical}. It is therefore plausible that a female has evolved to consume conspecific eggs at a lower rate, since there is a non-negligible chance that these conspecific eggs were oviposited by the female herself. This kin selection argument is speculative, but it would explain the consistency of priority effects across Tribolium competition experiments, despite significant differences between species and strains in many aspects of life history \citep{sokoloff1965productivity, park1965cannibalistic, karten1965genetic}.

Our microcosm experiments address long-standing topics in Tribolium research, such as the importance of demographic stochasticity vs. genetic founder effects, evolutionary explanations for priority effects, and dubious coexistence in Park et al.'s long-term experiments. We have also demonstrated that deterministic models can reliably predict competitive outcomes, even in small communities with substantial demographic stochasticity. The framing of determinism vs. stochasticity is particularly relevant due to an explosion of interest in fluctuation-mediated coexistence, contrasting with the long-standing tradition of using deterministic Lotka-Volterra models (and other, structurally similar models). Detailed case studies of natural communities and microcosms are crucial for establishing the relevance of stochasticity, with the ultimate goal of validating a general methodology for studying ecological coexistence.

\section{Acknowledgements}  \label{Acknowledgements} 

We thank the Hastings Beetle Lab team of undergraduates (Alison Nguyen, Jessica Au, Ivan Beas, Zach Cornejo, Natalie Flores, Shannon Mcgraw, Kenzie Pollard, Charlotte Rappel, Harsitha Sakhamuri, Mitali Singh, Alyssa Somers, Katie Somers, Jia Wang, and Emily Wood), at the University of California Davis, for assisting with data collection. TAD was supported by the National Science Foundation (NSF-DEB-2420769 and NSF-DEB-2017826).

\section{Data and code availability statement} \label{Data and code availability statement}
All code and data are available on Zenodo (\url{https://zenodo.org/records/13629641}).

\begin{appendices}
\counterwithin{figure}{section}

\section{Supplementary figures} \label{Supplementary figures}

\begin{figure}[H]
\centering
\makebox{\includegraphics[scale = 0.85]{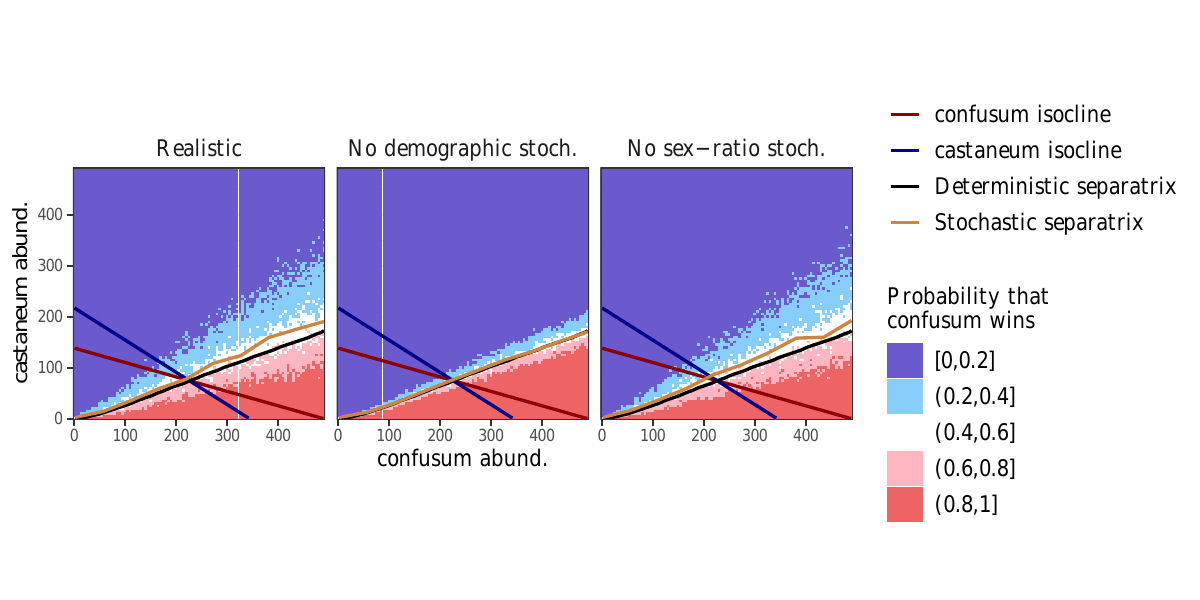}}
\caption{Demographic stochasticity is responsible for most of the competitive indeterminacy. Stochastic phase plane diagrams are shown for various simulation scenarios. Competitive indeterminacy, which may be visualized here as the width of the light-colored funnel (probability that \textit{confusum} $\in (0.2, 0.8]$), is greatly reduced when demographic stochasticity is eliminated.}
\label{fig:stoch_phase_scenarios}
\end{figure}

\begin{figure}[H]
\centering
\makebox{\includegraphics[scale = 1]{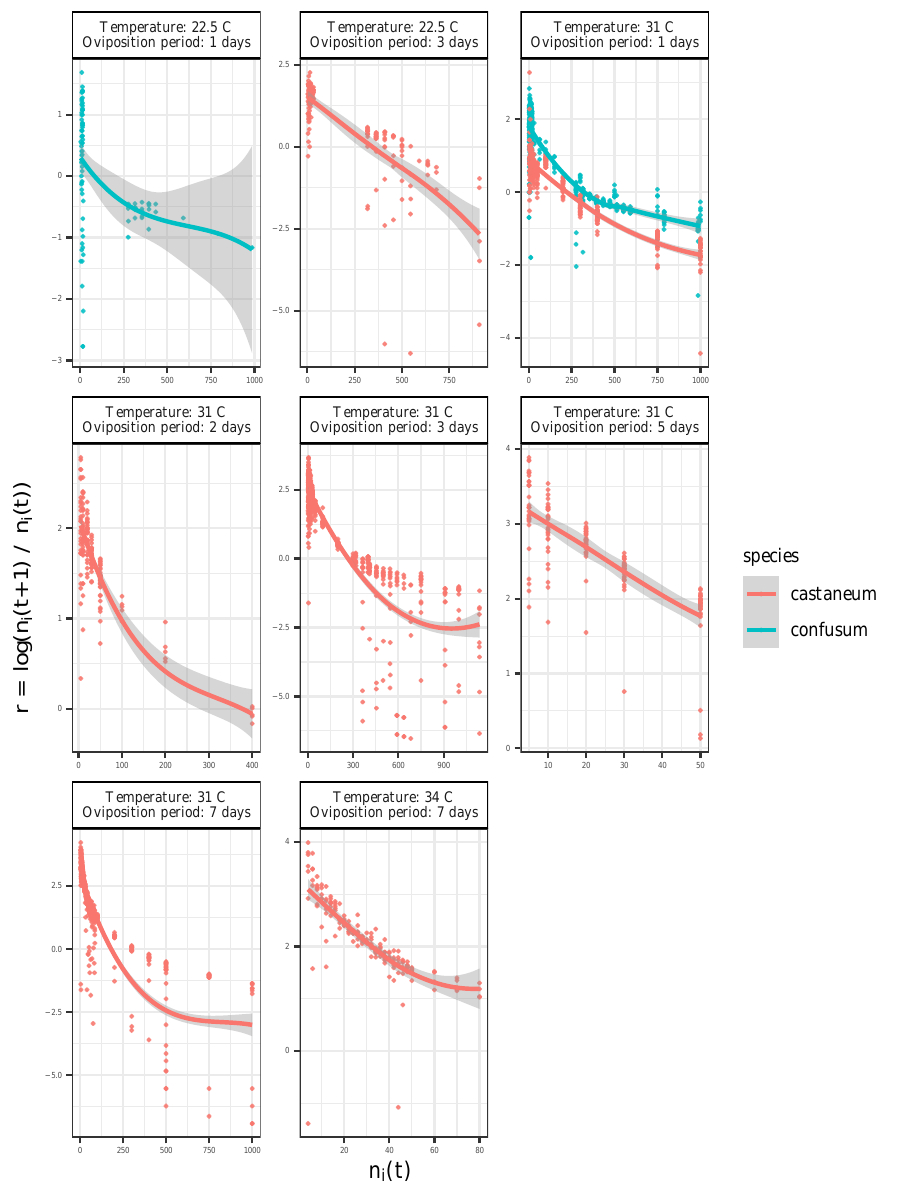}}
j\caption{Per capita growth rates, $r$, are a concave-up function of the population abundance $n_i(t)$. Different facets show different experimental conditions. Curves are generated by loess with span = 1.}
\label{fig:growth_rates}
\end{figure}

\counterwithin{table}{section}
\end{appendices}
\bibliographystyle{apalike}
\bibliography{refs.bib}

\end{document}